\documentclass[twoside,twocolumn]{article}

\usepackage[sc]{mathpazo} 
\usepackage[T1]{fontenc} 

\usepackage[english]{babel} 

\usepackage[hmarginratio=1:1, vmarginratio=1:1, top=15mm, left=18mm, columnsep=10pt]{geometry} 
\usepackage[small,labelfont=bf,up,textfont=it,up]{caption} 
\usepackage{booktabs} 
\usepackage{xcolor}
\usepackage{soul}

\newcommand{\ie}{\emph{i.e.}}

\newcommand{\kBT}{{k_\te{B} T}}

\newcommand{\rhos}{\rho_\te{s}}

\newcommand{\te}[1]{\mathrm{#1}}

\newcommand{\ev}[1]{\left\langle #1 \right\rangle}

\usepackage{abstract} 

\usepackage{titlesec} 
\renewcommand\thesection{\Roman{section}} 
\renewcommand\thesubsection{\roman{subsection}} 
\titleformat{\section}[block]{\large\scshape\centering}{\thesection.}{1em}{} 
\titleformat{\subsection}[block]{\large}{\thesubsection.}{1em}{} 

\usepackage{fancyhdr} 
\usepackage{authblk}

\author[1,2,3]{Joscha~Mecke}
\author[1]{Yongxiang~Gao\thanks{Email: yongxiang.gao@szu.edu.cn}}
\author[3]{Gerhard~Gompper}
\author[3]{Marisol~Ripoll\thanks{Email: m.ripoll@fz-juelich.de} }

\affil[1]{\small{Institute for Advanced Study, Shenzhen University, 518060 Shenzhen, P.R. China}}
  \affil[2]{\small{College of Physics and Optoelectronic Engineering, Shenzhen University, 518060 Shenzhen, P.R. China}}
  \affil[3]{\small{Theoretical Physics of Living Matter, Institute of Advanced Simulation,}\protect\\ \small{Forschungszentrum Jülich, 52425 Jülich, Germany}}

\usepackage{titling} 

\usepackage{hyperref} 

\usepackage{graphicx}
\usepackage{subcaption}

\usepackage{amsmath}
\usepackage{amssymb}

\usepackage[superscript,biblabel]{cite}

\usepackage{bm}

\pretitle{\begin{center}\huge\bfseries} 
\posttitle{\end{center}} 
\title{Chiral active systems near a substrate: Emergent damping length controlled by fluid friction}
\date{July 3, 2024} 
\renewcommand{\maketitlehookd}{%
\begin{abstract}
  \noindent Chiral active fluids show the emergence of a turbulent behavior characterized by multiple dynamic vortices whose maximum size is specific for each experimental system. This is in contrast to  hydrodynamic simulations in which the size of vortices is only limited by the system size. 
We propose and develop an approach to model the effect of friction close to a surface in a particle based hydrodynamic simulation method in two dimensions,
in which the friction coefficient can be related to the system parameters and to the emergence of a damping length. 
This length limits the size of the emergent vortices, and influences other relevant system properties such as the actuated velocity, rotational diffusion, or the cutoff of the energy spectra. 
Comparison of simulation and experimental results show a good agreement which demonstrates the predictive capabilities of the approach, which can be applied to a wider class of quasi-two-dimensional systems with friction.
\end{abstract}
}

\begin{document}

\maketitle

                              
\section*{\label{sec:level1}Introduction}

Friction refers to the dissipation of energy given the relative motion of two different elements, which can occur between two solid surfaces, two fluids, and also between a liquid and a solid substrate. Complex fluids, such as solutions of colloidal, polymeric, or biological components are mostly investigated in confinement or in the vicinity of a solid substrate, whose specific effect is frequently not taken into account. 
Active matter systems are characterized by a nonequilibrium motion of microscopic units~\cite{gompper20202020}, such as swimming bacteria~\cite{baskaran2009statistical} or algae~\cite{drescher2009dancing}, cancer cell motility~\cite{grosser2021cell, gottheil2023unjamming}, or optically controlled synthetic colloidal particles~\cite{soker2021landscape}. %
Ensembles of such active units display a rich variety of macroscopic collective behaviours such as swarming~\cite{wagner2021collective}, motility induced phase separation~\cite{roca2022clustering}, and an extensive number of further structures and collective motions~\cite{negi2022emergent}, which can occur in bulk or confinement. %
In the case of active matter, where components have an intrinsic or externally activated motion, the friction might be of special relevance, since the relative motion of the active components and the passive substrate is intrinsically larger. 
Examples where substrate friction has explicitly shown to play an essential role are found in propagation mechanisms such as cell migration~\cite{vazquez2022substrate}, in the dynamics of active nematic systems~\cite{thampi2014active} or in turbulent flows~\cite{boffetta2005effects}.

Hydrodynamic interactions (HI) play an important role in many active matter suspensions~\cite{elgeti2015physics}, in particular in the presence of an interface or the proximity of walls, where HI can coexist with substrate friction. Due to HI with the walls, swimming bacteria experience an increased probability of staying and eventually performing circular trajectories close to the walls~\cite{berke2008microorganisms,mousavi2020bacteria},  and spermatozoa to self-assemble into a lattice of vortices of equal direction of circulation at planar surfaces~\cite{riedel2005self}.
Hydrodynamic interactions are also responsible for the creation of large-scale vortices in chiral active suspensions~\cite{mecke2023simultaneous}. 
Chiral active matter is composed of units that spin or perform a circular motion around a fixed axis~\cite{lowen2016chirality, bechinger2016crowded,furthauer2012active} whose hydrodynamic and steric interactions give rise to rich cooperative effects such as the emergence of multiscale vortices~\cite{petroff2015rotating,zhang2020reconfigurable,mecke2023simultaneous,mecke2024emergent}. %
The so-called odd viscosity in chiral active fluids is an antisymmetric contribution to the viscosity tensor which acts perpendicular to applied shear stresses and leads to correlations between density and vorticity, which vanish in isotropic three-dimensional systems~\cite{mecke2023simultaneous,fruchart2023odd,souslov2019topological, banerjee2022hydrodynamic,avron1998odd}.
Most of chiral systems are by construction composed of rotating colloids trapped at interfaces or substrates, where the dynamics is restricted to a two-dimensional plane.
Experimental observations have shown the emergence of a relevant length scale limiting the vortex size which is system dependent and which has not yet been rigorously related to an underlying mechanism. 
 
From a computational perspective, there are two main approaches to model a complex hydrodynamic fluid at a planar interface or close to a planar substrate. The first approach is to model the complete three-dimensional system where the boundary conditions at the confining walls or interfaces have to be specified~\cite{menzel2016dynamical,hosaka2021nonreciprocal, hosaka2023surfer, jia2022interface}. These conditions range from no-slip boundary conditions at a wall to a continuous coupling with another fluid, and therefore specific to each experimental conditions. 
In systems where the relevant suspension dynamics takes place in two dimensions, the consideration of the third dimension leads to a considerable increase in analytical or computational effort~\cite{theers2016modeling,qi2022emergence}, which  ultimately poses a critical limit for the size of the systems under study. %
The second approach is to simply model a two-dimensional suspension in which the colloidal degrees of freedom are restricted to a plane. This enormously reduces the computational effort, and allows for a clear interpretation of the system relevant dynamics. However, since hydrodynamic interactions are very long ranged in two dimensions, divergent or other unphysical behaviour might occur
such as the Stokes or the Jeffery paradoxes~\cite{jeffery1922rotation}. The Stokes paradox states that there can be no creeping flow of a fluid around a disk in two dimensions, while the Jeffery paradox discloses that no solution to the Stokes equation can simultaneously fulfil the no-slip boundary condition caused by two equally sized disks rotating with equal angular velocities and opposite directions in a resting and infinitely large two-dimensional fluid.
This means that given two close rotors in a two-dimensional setup, if the flow velocity is never completely damped, a finite flow velocity will remain at infinity~\cite{ueda2003cylinders}. 
In realistic setups, two effects resolve this problem. 
The induced two-dimensional flows can partially escape into the third dimension enforcing that the fluid rests at infinity. Alternatively, solvent and suspended particles can be considered to have a non-negligible friction with a substrate, dissipating therefore energy which has then a similar effect, namely that the fluid rests at infinity.

In this work, we consider a well-established mesoscopic hydrodynamic algorithm, multiparticle collision dynamics (MPC)~\cite{gompper2009multi} (see the "Methods" section) and propose an extension to account for the momentum dissipation into a substrate. This is achieved by the exchange of momentum with a reservoir, which here takes the form of virtual particles. 
Our model is in excellent agreement with the expected 
Brinkman equation~\cite{brinkman1949viscous} which considers hydrodynamics with a linear damping term, with a friction coefficient which nicely matches the analytic prediction here performed in terms of simulation parameters. 

%
The method is then applied to a chiral active fluid, proving that the friction between substrate and fluid introduces a cutoff distance for the hydrodynamic interactions. %
As a first consequence, this means that in general the substrate friction is able to eliminate the divergences of the two-dimensional HI, such as the Jeffery paradox for a system of couple counter-rotating colloids~\cite{mecke2023birotor}. 
Furthermore, the consideration of the substrate friction clearly shows to be origin of the emergence of a typical damping length in chiral active systems, for which this length is strongly related to the most relevant system properties, such as vortex size, actuated average particle velocity, rotational diffusion, or the  small wavevectors cutoff of the power spectra. The model parameters can be then tuned such that simulations accurately reproduce the observed experimental average behavior.

%
\section*{Results}
\subsection*{\label{sec:model:substrate}Modelling substrate friction}

\begin{figure*}[ht!]
    \centering
    \includegraphics[width=0.9\textwidth]{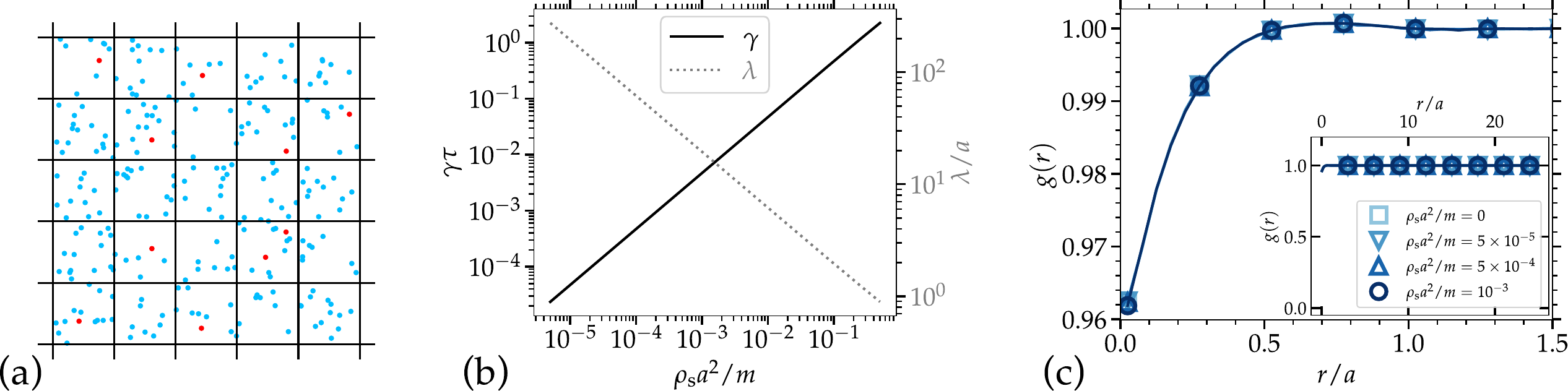}
	\caption{\label{fig:intro}
    {\bf Substrate friction for a two dimensional MPC solvent.}
	a)~Illustration of a collision step with MPC solvent particles (in blue) interacting with a set of random virtual substrate particles (in red).  
    b)~Friction coefficient $\gamma$, and the related decay length $\lambda$ values in terms of the virtual particles number density $\rhos$ as given by  Eq.~(\ref{eq:gamma-short}) and Eq.~(\ref{eq:lambda}), respectively. 
    c)~Radial distribution function of particles for simulations of a fluid in equilibrium with different densities of substrate virtual particles $\rhos$. Inset is a zoom out, showing that substrate virtual particles do not modify the fluid structure. %
	}
\end{figure*}
Substrate friction refers to the dissipation of energy and momentum when a fluid interacts with an interface, which can be air or a solid wall at solvent-air or solvent-substrate interfaces. 
The key idea used here is that influence of friction is similar to the interaction of the solvent with a momentum reservoir with zero mean momentum. 
In the case of a fluid at rest, the interaction with the reservoir conserves energy and momentum, and simply providing a certain degree of memory lost.
In the case of a fluid with some average velocity, the interaction with the reservoir implies that the dissipated energy is proportional to the fluid velocity. The proportionality factor, which we refer as $\gamma$,  is then the friction coefficient. 

In the here considered MPC method, the fluid is described by a number of point particles with alternating streaming steps, where particles freely propagate, and collision steps (see the "Methods" section).
Solvent interactions are reduced to the momentum interchange among particles in the same collision cell in every collision step, where each particle's velocity $\bm{v}_i(t)$ relative to the cell centre-of-mass velocity $\bm{v}_\zeta(t)$ is rotated by an angle of $\pm\vartheta$. 
We modify the MPC collision step by considering additional randomly distributed virtual particles, as illustrated in Fig.~\ref{fig:intro}a.
In general, each collision box is now populated by the $N_\zeta$ MPC fluid particles  and $M_\zeta$ additional virtual particles.
The cell centre-of-mass velocity is then, 
\begin{align}
    \bm{v}_\zeta(t) =  \frac{\sum_{i}^{N_\zeta} \bm{v}_i(t) + M_\zeta\hat{\bm{v}}}{N_\zeta + M_\zeta}\,,
\end{align}
The velocities $\hat{\bm{v}}$ of the virtual particles are drawn randomly from a Maxwell-Boltzmann distribution, with average $\ev{\hat{\bm{v}}}=0$ and variance $\ev{\hat{v}^2}=\kBT/m$.  
In that way, momentum can be transferred from the fluid to the substrate in each collision step. This constitutes a minor adjustment in the simulation code and only marginally increases the computational effort. 
The transfer of momentum with the average density of the virtual particles $\rhos$. 
This density $\rhos$  is an adjustable parameter of the simulations, and can be related to the friction coefficient $\gamma$.  

Although substrate friction was not described in this way before, the interaction of MPC and virtual particles has been previously considered to improve stick boundary conditions with planar walls and with spherical colloids~\cite{lamura2001multi,gotze2011flow,gotze2011dynamic}, and also to describe the so-called random  MPC solvent~\cite{ripoll07epje}, employed to quantify the importance of hydrodynamic interactions in a number of systems~\cite{liebetreu18}.
The method here proposed can also be easily generalised to a number of different geometries and used both in two- and three-dimensional systems. 
We focus in the two-dimensional case, where the approach is particularly useful since the very long range of hydrodynamic interactions in two dimensions decreases when friction is considered. 
When applied to a two-dimensional problem, an alternative to this model with virtual friction particles is to consider a quasi-two dimensional fluid, \ie, a three-dimensional system strongly confined in between parallel walls with stick boundary conditions~\cite{theers2016modeling}. This might have advantages in specific problems, but  computationally is significantly more costly. 
On the other hand, note that the inclusion of each virtual particle can also be understood as the interaction with a rough surface element, obstacle or a pore, such that the overall dynamics is supposed to be similar to that of a porous media. 

The resulting friction coefficient $\gamma$ in a MPC fluid is determined by the momentum transfer due to the presence of the virtual particles. We performed an analytical estimation of this coefficient (see the "Methods" section), which for a rotation angle of $\vartheta = \pi/2$ the relation simplifies to
\begin{align}
\label{eq:gamma-short}
    \gamma = \frac{1}{h} \frac{\rhos}{\rho + \rhos} \left[ 1- \frac{1}{2(\rho - 1)} \right]\,,
\end{align}
where $h$ is the time between collisions and $\rho$ the average number of particle in a collision cell.
This result implies that the friction coefficient is directly proportional to the relative number density of virtual particles, which in the limit of $\rho \gg \rhos$ implies that $\gamma \propto \rhos/h$.


\subsection*{Solvent damping length}
Related with the friction coefficient and the solvent kinematic viscosity $\nu =\eta/\rho$, a typical solvent damping length can be introduced as,
\begin{equation}
    \lambda = \sqrt{\frac{\nu}{\gamma}}.
\label{eq:lambda}
\end{equation}x
This length determines the extend of the fluid momentum decay. This can be understood as the length an external perturbation to the fluid takes to decay in an infinitely large system without any further interaction. 
In practical applications, this length $\lambda$ determines a system typical behaviour, which acts in combination with the other relevant system parameters.

Very small friction can make this length much larger than the system size, which implies that friction has very little effect. However, very large friction values render this length as small as the collision box size, or even smaller, which would lead to complete decorrelation, a limit in which hydrodynamic interactions do not play any role.  We are therefore typically interested in an intermediate range of damping lengths.  

For the simple MPC fluid, both the friction and the decay length can be calculated in terms of $\rhos$ with Eqs.~(\ref{eq:gamma-short}) and~(\ref{eq:lambda}). 
Fig.~\ref{fig:intro}b shows that already relatively small values of $\rhos$ result in decay lengths of the order of the collision box, and that for example one single virtual particle every $1000$ collision boxes is related to a decay length of $25a$. 

\subsection*{Fluid in equilibrium}

We first consider the trivial case of a two-dimensional fluid at rest, simulated with periodic boundary conditions.  
In MPC fluids, the density pair correlation function $g(r)$ is in principle expected to be the uniform distribution of the ideal gas~\cite{ripoll2005dynamic}. 
Small deviations from this behaviour can be traced back to the breakdown of molecular-chaos, \ie, fluid particles may perform several collision on short time scales with the same neighbours, resulting in slightly correlated dynamics of the MPC particles~\cite{ripoll2005dynamic,noguchi2008transport}. 
The radial distribution function of the solvent particles for fluids with various values of $\rhos$ is measured in simulations and shown in Fig.~\ref{fig:intro}c. The results verify that the introduction of substrate virtual particles has no influence on the effective potential between the fluid particles independent of the density $\rhos$. Accordingly, other fluid properties arising from effective intermolecular potentials between the MPC particles like the compressibility or viscosity are unaffected from the introduction of the substrate friction.

\subsection*{Circular Couette flow}

\begin{figure*}[ht]
	\centering           
        \includegraphics[width=0.65\textwidth]{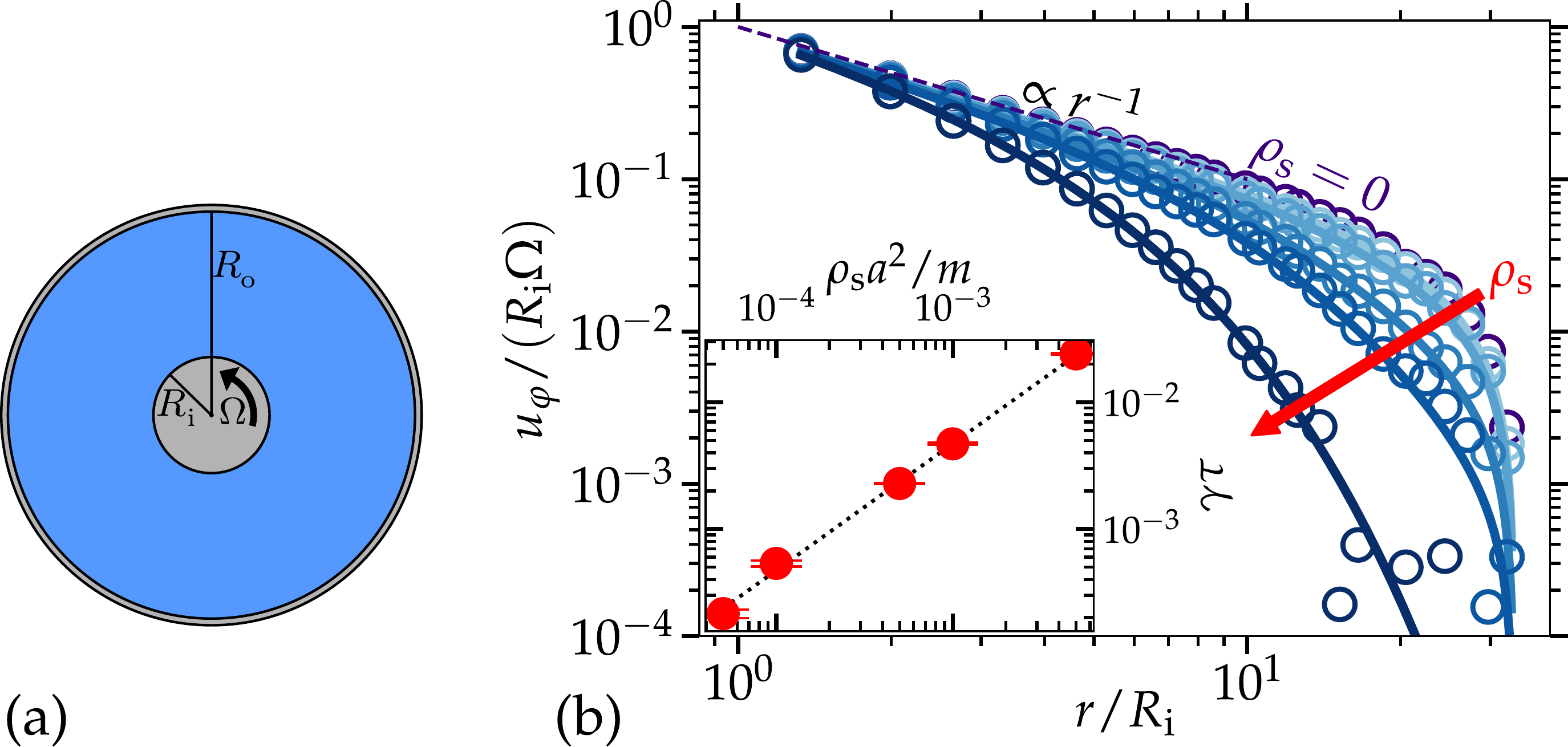}
        \caption{
        {\bf Circular Couetter flow.}
        a)~Sketch  of the simulated Couette flow rheometer. The fluid is confined between concentric cylinders, the inner cylinder has a radius $R_\te{i}$ and constant rotation velocity $\Omega$, the outer cylinder has a radius $R_\te{o}$, and is fixed. 
        b)~Fluid velocity as a function of the 
        radius for fluids with increasing $\rho_\te{s}10^3 a^2/m = 0.0, 0.05, 0.1, 0.5, 1.0, 5.0$. Purple line corresponds to $\rhos=0$ and increasing blue darkness correspond to increasing friction, as indicated by the arrow. Symbols are simulations averaged velocities, and solid
        lines correspond to Eq.~\eqref{eq:rad-fric-sol}, with $\gamma$ as a fitting parameter. 
        In the inset the $\gamma$ values obtained from simulations (bullets) are displayed as a function of the input $\rhos$ values, together with the prediction from Eq.~\eqref{eq:gamma-short} (dotted line), showing a very good agreement. 
        Errorbars are extracted from the least-square fits and are smaller than the symbol size. %
	}
	\label{fig:couette}
\end{figure*}

The effect of substrate friction is tested in the geometry of circular Couette flow. 
The fluid is restricted in the area enclosed by two concentric cylinders. The inner circle, with radius $R_\te{i}$, is rotating at angular frequency $\Omega$, and the outer cylinder, with $R_\te{o} > R_\te{i}$, does not move, as schematically shown in Fig.~\ref{fig:couette}a. Both surfaces have no-slip boundary conditions~\cite{gotze2011flow}, such that the fluid velocity continuously varies in between the surfaces. The functional form of this variation is not directly imposed in the simulations, but simply provided by the MPC interactions. The hydrodynamic behaviour is expected to agree with Stokes equation with the inclusion of the friction term, this is the Brinkman equation in Eq.~\eqref{eq:stokes-friction}.

Given the radial symmetry of the system, the velocity has only a non-vanishing tangential velocity, such that the velocity can be considered in polar coordinates $\bm{u} = u_\varphi(r) \hat{e}_\varphi$. With this the Brinkman equation takes the form of the Bessel equation (see the "Methods" section), such that by considering the imposed velocity at the two boundary cylinders, namely  
$u_\varphi(R_\te{i}) = R_\te{i}\Omega$ and $u_\varphi(R_\te{o}) = 0$, %
the flow field can be calculated as, 
\begin{equation}
	\label{eq:rad-fric-sol}
	u_\varphi(r) = R_\te{i}\Omega \frac{ J_1(-\widetilde{R}_\te{o}) Y_1(-\widetilde{r}) - Y_1(-\widetilde{R}_\te{o}) J_1(-\widetilde{r}) }{J_1(- \widetilde{R}_\te{o}) Y_1(-\widetilde{R}_\te{i}) - Y_1(-\widetilde{R}_\te{o}) J_1(-\widetilde{R}_\te{i})},
\end{equation}
with the variables change $\widetilde{x}\equiv i x/\lambda$, and with $J_1(x)$ and $Y_1(x)$ the Bessel functions of the first and second kind, respectively~\cite{akhmedova2019selected}.
In the limit of vanishing friction ($\lambda \to \infty$), the flow profile attains the result of the Stokes equation,
\begin{align} \label{eq:rad-sol}
	u_\varphi(r) = - \frac{\Omega R_\te{i}^2}{R_\te{o}^2 - R_\te{i}^2}r + \frac{\Omega R_\te{i}^2 R_\te{o}^2}{R_\te{o}^2 - R_\te{i}^2} \frac{1}{r}\,,
\end{align}
which decays as $u_\varphi \propto r^{-1}$ at distances close to the inner circle $r/R_\te{i}  \gtrsim  1$, 
and vanishes at the outer circle,  as dictated by the no-slip boundary.  %

Simulations with an MPC fluid in two dimensions with the geometry in Fig.~\ref{fig:couette}a are performed for $R_\te{i}=3a$ and $R_\te{o}=90a$, and various values of $\rhos$.  
A stationary flow profile is reached, and flow profiles are shown in Fig.~\ref{fig:couette}b, together with the analytical prediction in Eq.~\eqref{eq:rad-fric-sol}, 
whose excellent agreement proves that our approach fulfills the Brinkman equation. %
In the frictionless case, Eq.~\eqref{eq:rad-sol} is used, and for the cases with friction, the decay length $\lambda$ is treated as a fit parameter according to Eq.~\eqref{eq:rad-fric-sol}. This provides a direct measurement of the actual damping length and with Eq.~(\ref{eq:lambda}) of the fluid friction coefficient. 
The inset of Fig.~\ref{fig:couette}b shows a comparison of the substrate friction coefficient measured in simulations and the prediction from Eq.~(\ref{eq:gamma-short}), which demonstrates an excellent agreement. 
Note that the smallest non-vanishing friction here employed is $\gamma \tau=0.0002$ which corresponds to $\lambda/R_\te{i}=33$, this is larger than the system size, such that the flow profile is almost the same as the frictionless case. Increasing the friction increases the deviation from the Stokes solution in Eq.~\eqref{eq:rad-sol}, and for the largest simulated friction, $\gamma \tau=0.02$, the deviation is significant at the corresponding decay length $\lambda/R_\te{i}=3.3$. 

\subsection*{Chiral active fluids}

\begin{figure}[h]
    \centering
    \includegraphics[width=.8\columnwidth]{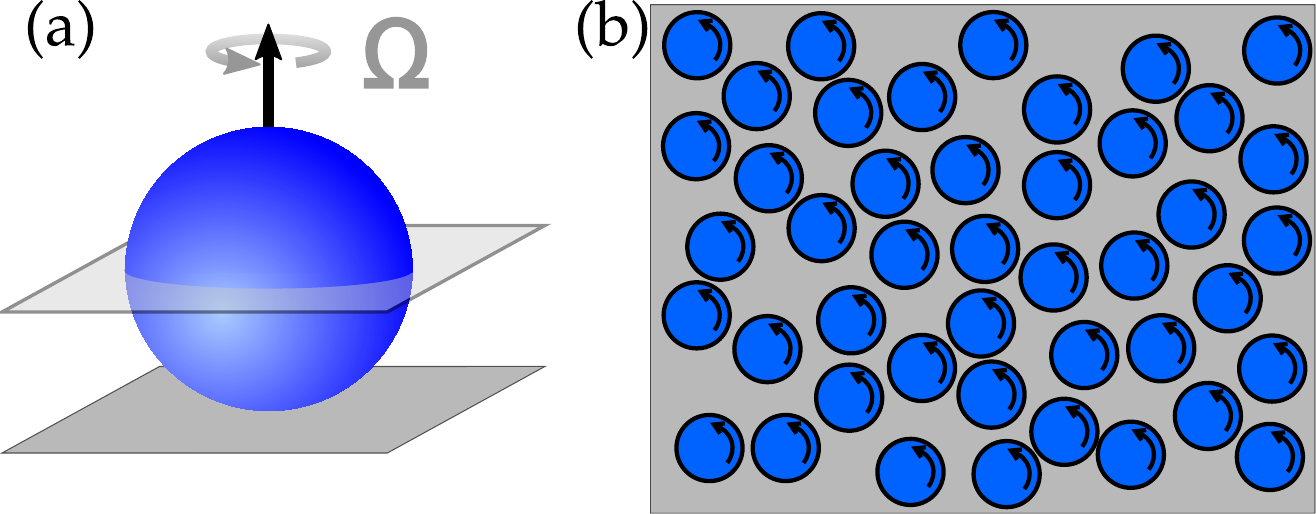}
    \caption{
        {\bf Sketch of a chiral active fluid setup on a substrate related to Refs.~\cite{soni2019odd, bililign2022motile, mecke2023simultaneous}}. %
        (a) The colloids are sedimented to the substrate (dark plane) such that the colloidal translational dynamics is restricted to a plane. The colloids' chiral activity stems from an inherent particle rotation around the substrate normal. %
        (b) The fixed axis of chirality perpendicular to the substrate and the two-dimensional centre of mass dynamics of the colloids allow to follow the dynamics of the colloids only in the 2D plane, as seen from above.
    }
    \label{fig:sketch}
\end{figure}

We consider a chiral active fluid consisting of a suspension of colloidal particles sedimented to a substrate, which due to an intrinsic property, such as a dipolar moment, or a permanent magnetic moment, rotate under the influence of a rotating external field~\cite{mecke2023simultaneous,soni2019odd, massana2021arrested,bililign2022motile,zhang2020reconfigurable} (see Fig.~\ref{fig:sketch}). 
Due to the no-slip boundary condition on the colloid surface, each rotating colloid drags a rotational flow in the surrounding fluid decaying like $r^{-1}$ , affecting neighbouring colloids. 
Given the system symmetry, the hydrodynamic equations of motion of such systems can be regarded as essentially two-dimensional~\cite{soni2019odd}. 

We perform simulations of colloidal discs with diameter~$\sigma$ rotating at a constant angular velocity $\Omega$ in a two-dimensional MPC fluid. 
The coupling between colloidal and solvent degrees of freedom is achieved by regarding the colloids as moving impenetrable no-slip boundaries that exchange linear and angular momentum with the fluid during collision and streaming steps~\cite{mecke2023simultaneous,mecke2023birotor, gotze2011dynamic,gotze2011flow}.

\bigskip
\noindent{\bf Vorticity field. \ }
The centre-of-mass of isolated rotors display a purely diffusive motion without any active displacement.  
When colloids are not isolated, the created rotational flow fields drag nearby colloids. 
For two rotors, this leads to a pairwise orbital rotation of the same sign as the imposed colloid rotation around the centre of mass of two rotors~\cite{mecke2023simultaneous, fily2012cooperative, climent2006dynamic}. %
In a rotor ensemble, the many-body hydrodynamic interactions lead to the formation of a cascade of vortices, \ie, vortices of different sizes form and the larger vortices themselves consist of smaller vortices and eventually of the orbital rotation of rotor pairs~\cite{mecke2023simultaneous}. 
The space between vortices of the same circulation is occupied by a vortex flow of opposite circulation, as a result of the continuity of flows, and the collective dynamics is reminiscent of active turbulence. %

\begin{figure*}
	\centering
	\includegraphics[width=\textwidth]{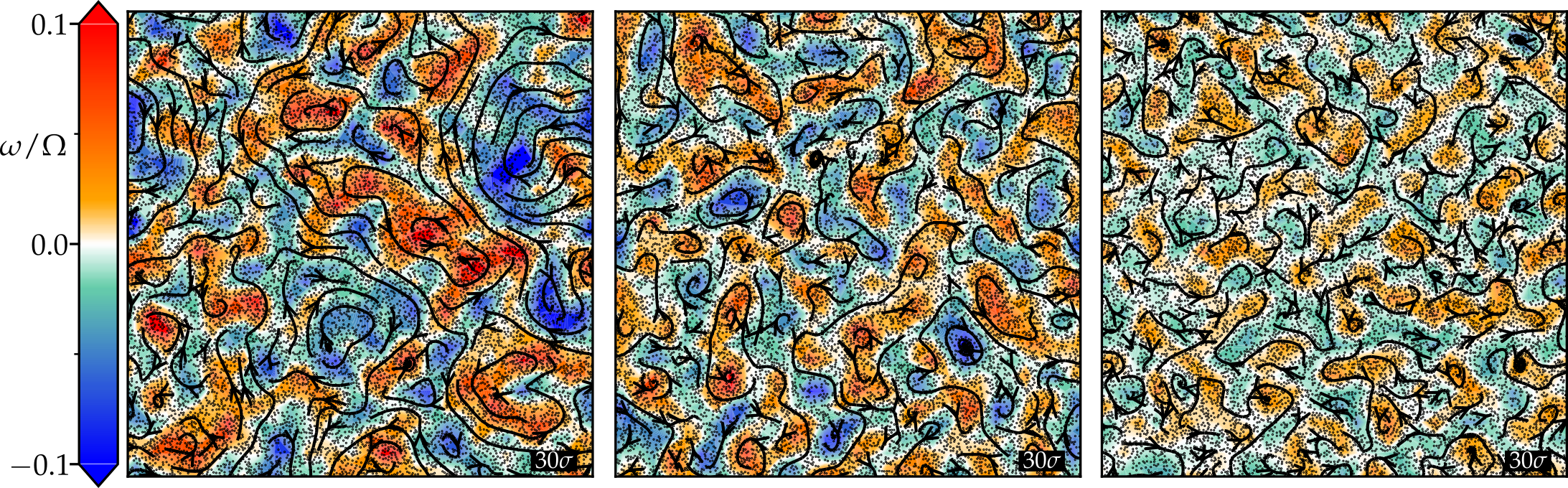}
	\caption{
        {\bf Streamlines and vorticity fields of a chiral active fluid.}
		Coarse-grained vorticity fields with superimposed streamlines and underlying colloid positions, for simulations with substrate friction coefficients increasing from left to right, 
   $(\gamma \tau) 10^3 = 0, 0.2, 2.4$ and $L=300\sigma$, showing the decrease of both, the vorticity magnitude and the vortex size with increasing $\gamma$. 
	}
	\label{fig:vorticity-density-collective}
\end{figure*}

The coarse-grained velocity field, $\bm{v}(r)$, is calculated from the colloids instantaneous displacement, which vary with time and position. From such velocity field the streamlines of the collective colloid dynamic can be calculated, as shown in Fig.~\ref{fig:vorticity-density-collective} together with the corresponding colloid vorticity fields $\omega = \partial_x v_y - \partial_y v_x$.
Both, the streamlines and the vorticity fields show the presence of collective vortices of multiple sizes. 
Note that although the system is heterogeneous, there is no preferential orientation such that the system can be consider to be isotropic. 

In the absence of substrate friction, the size of the vortices is only limited by the size of the simulation box, such that vortices ranging over more than half of the domain might emerge, as can be seen in Fig.~\ref{fig:vorticity-density-collective}a.
In the presence of substrate friction, the flows induced by each rotor experience an exponential cutoff with length scale controlled by the strength of the substrate friction, analogous to the attenuation of the created flows in circular Couette flow. %
In a concentrated system, the decrease of the flow intensity translates into smaller values of the induced vorticities, $\ev{\omega^2}$, and the fastest spatial decay into the decrease of the average vortex size. %
Simulations with increasing values of the substrate friction in Fig.~\ref{fig:vorticity-density-collective}b and Fig.~\ref{fig:vorticity-density-collective}c clearly show the decrease of both, the vorticity magnitude and vortex size. 
The size of these vortices can be estimated in various ways, which do not necessarily provide the same value.   
     
\begin{figure}[!ht]
		\includegraphics[width=0.48\textwidth]{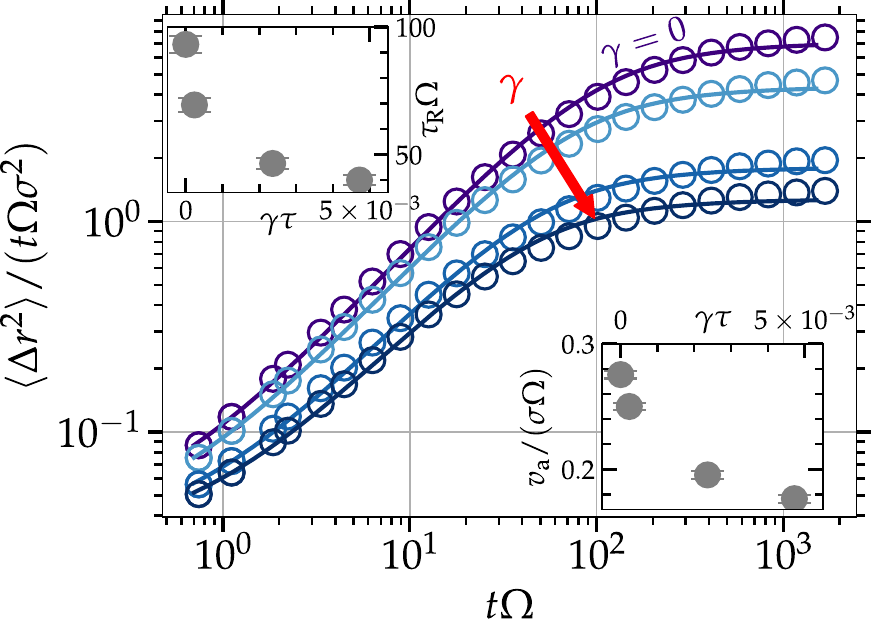}
	\caption{
        {\bf Average translation properties of a chiral active fluid.}
		Mean-square displacement in a system with friction coefficients $(\gamma \tau) 10^3 = 0, 0.2, 2.4,4.7$ and $L=100\sigma$. 
        Purple line corresponds to $\gamma=0$ and increasing blue darkness correspond to increasing $\gamma$. 
        Symbols are the simulation averages and lines correspond to least-square fits to the active Brownian particle model~\cite{howse2007self}. %
        The arrow is a guide to the eye denoting increasing $\gamma$ values.
        Top and bottom insets: Rotational diffusion times and actuated velocity as a function of $\gamma$ obtained from $\ev{\Delta r ^2}$.  
	}
	\label{fig:msd_substrate}
\end{figure}

\bigskip
\noindent{\bf Particle trajectories. \ }
For a rotor ensemble, the actuated trajectory of each rotor constantly changes in between vortices of different sizes and locations, such that the averaged single rotor motion can be understood as an active Brownian trajectory~\cite{mecke2023simultaneous}. 
The averaged mean squared displacement is 
\begin{equation}
\ev{\Delta r ^2}=\ev{\left[r(t+t_0)-r(t_0)\right]^2}    
\end{equation}
with the angles indicating average over particles, initial times $t_0$, and ensembles. The values calculated from the simulation data are shown in Fig.~\ref{fig:msd_substrate} for various values of the friction coefficient.
The average actuated velocity and rotational diffusion timescale of single rotors is obtained by performing least-square fits of the mean-square displacements to the active Brownian particle model~\cite{howse2007self} and shown in the insets of 
Fig.~\ref{fig:msd_substrate}. The time-normalised mean-square displacement for the rotor system for different values of $\gamma$ in Fig.~\ref{fig:msd_substrate} reveals two influences of the substrate friction on the rotor trajectories. 
On the one hand, the flow attenuation due to substrate friction decreases the mutually actuated velocity such that the overall dynamics is slowed down. 
This implies that values of  $\ev{\Delta r ^2}$ decrease with friction already at short times, and that the average velocities decay with $\gamma$ (see bottom inset in Fig.~\ref{fig:msd_substrate}). %
On the other hand, since angular diffusion takes place on the timescale of the largest possible vortex, the reduction of the vortex size with increasing friction makes the rotational diffusion coefficient $\tau_\mathrm{R}$ to decrease with $\gamma$ (see top inset in Fig.~\ref{fig:msd_substrate}). This is also evident in the mean-squared displacement curves, since the decrease  with friction  is more prominent at large time than at short times. 

\begin{figure}[!ht]
		\includegraphics[width=0.48\textwidth]{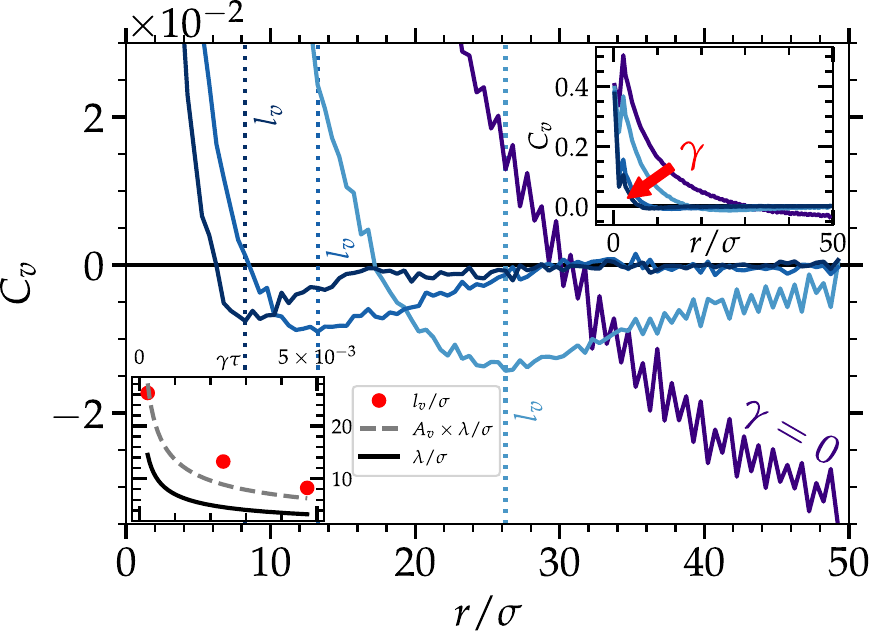}
	\caption{
        {\bf Vortex size of a chiral active fluid.}
		Velocity spatial correlation function in a system with $L=100\sigma$ with the same $\gamma$ values and color code as in Fig.~\ref{fig:msd_substrate}.
        Vertical dotted lines indicate the positions of the $C_v$ minima, called here $l_v$, for each $\gamma$ value. %
		Top inset: Zoom-out of $C_v$ with the arrow denoting increasing $\gamma$ values. 
        Bottom inset: Dependence of the measured $l_v$ with the friction $\gamma$, comparison with $\lambda$ values in Eq.~(\ref{eq:gamma-short}) (solid line), and fit following the same dependence, as $l_v = 2.3 \lambda$ (dashed line). 
	}
	\label{fig:vcf_substrate}
\end{figure}

\bigskip 
\noindent{\bf Velocity correlation functions.}
An estimate of the average vortex size can be obtained by evaluating the spatial correlation of the spatial coarse-grained velocity field, $\bm{v}(r)$, which is calculated as a time and ensemble average,
\begin{equation}
C_v(r) = \frac{\ev{ \bm{v}(r+r_0) \cdot \bm{v}(r_0)}}{2\kBT/m_\te{c}}. 
\end{equation}
Here the thermal energy is used as normalizing factor, with  $m_\te{c}$ the mass of each colloid. 
Simulation results for different values of the substrate friction are shown in Fig.~\ref{fig:vcf_substrate}. 
The correlation decays with increasing distance, which occurs as a result of the flow decaying as $r^{-1}$ induced around each rotating colloid. 
An exception is $C_v(\sigma)$, which is smaller than $C_v(2\sigma)$. This is due to the strong contribution of pairs of colloids rotating around each other, which have therefore opposite directions.  

The correlation becomes negative at intermediate distances $r$ due to the existence of areas of both positive and negative vorticity, as shown in Fig.~\ref{fig:vcf_substrate} for all values of the employed friction. The position of the first minimum, denoted here as $l_v$, can then be regarded as an estimate to the average vortex size, since it relates to the average centre-to-centre distance of neighbouring couter-rotating vortices. 
%
For a system without substrate friction,  $l_v$ is located at half the system size, $L/2$. This means that, on average, the system dynamics is dominated by adjacent vortices of opposite circulation ranging over the whole system, which accommodate other large vortices due to periodic boundary conditions, and also smaller internal  vortices. %
With increasing  values of the substrate friction, 
the attenuation increases and the largest vortices decrease in size such that $l_v$ is shifted to shorter lengths, as shown in Fig.~\ref{fig:vcf_substrate}. 
The value of $l_\te{v}$ does not coincide with the value of damping length $\lambda$, but  the functional dependence with the friction $\gamma$ is the same, as can be seen in an inset of Fig.~\ref{fig:vcf_substrate}. The two values can be considered to be directly related by a correction factor $A_v$, with $l_v = A_v \lambda$ and $A_v=2.3$. 
Note also that with substrate friction and for long enough distances, the velocity correlation eventually vanishes, 
such that vortices or colloid flows distanced by $r \gg l_v$ are uncorrelated, as can be seen in Fig.~\ref{fig:vcf_substrate}.%

\bigskip
\noindent{\bf Energy spectra. \ }
An alternative approach to quantify the vortex size can be obtained from the energy spectra, which quantify the kinetic energy stored at different length scales.  
The translation energy spectrum can be obtained from the Fourier transformation of the velocity correlation function~\cite{landau1987fluid}, yielding
\begin{align}
    \label{eq:ennergy-spectrum}
    E_q = \frac{q}{4\pi} \ev{ \hat{\bm{v}}_q \hat{\bm{v}}_{-q} }\,.
\end{align}
We employ the two-dimensional discretised version of Eq.~\eqref{eq:ennergy-spectrum}, where the sum of the power spectral density is calculated from the colloid velocity field $\bm{v}$ over annular equal-$q$ shells~\cite{mecke2023simultaneous}.
\begin{align}
    E_q = \frac{1}{8\pi\Delta q} \sum_{q-\Delta q < k < q + \Delta q} \ev{ \hat{\bm{v}}_q \hat{\bm{v}}_{-q} },
\end{align}
where the wave vector $q$ is by construction limited by the colloid size as smallest relevant length scale, $q_\te{max} \simeq 2\pi/\sigma$, and the system size as  the largest relevant length scale, $q_\te{min} \simeq 2\pi/L$. %

\begin{figure}[!ht]
	\includegraphics[width=0.48\textwidth]{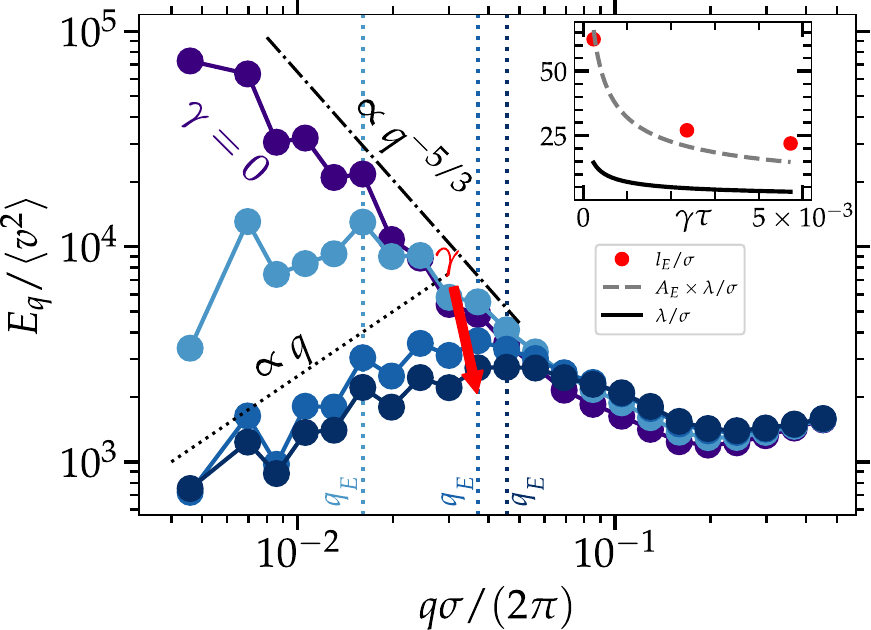}
	\caption{
        {\bf Energy spectra of a chiral active fluid.}
		$E_q$ in a system with $L=300\sigma$ with the same $\gamma$ values and color code as in Fig.~\ref{fig:msd_substrate}. %
        Vertical dotted lines indicate the positions of the $E_q$ maxima, defined here as $q_{_E}$. Dashed-dotted lines are a guide to the eye for the self-similar scaling behavior  $q^{-5/3}$, and the uncorrelated behavior $q^{-5/3}$, when the friction dominates. %
        Inset: Dependence of the measured $l_{_E}$ with the friction $\gamma$, comparison with $\lambda$ values in Eq.~(\ref{eq:gamma-short}) (solid line), and fit to the same dependence $l_{_E} = 4.6 \lambda$ (dashed line). 
}
	\label{fig:Eq}
\end{figure}
 The energy spectra of the rotor ensemble for simulations with different values for the substrate friction is shown in Fig.~\ref{fig:Eq}. %
While the mere rotation of the colloids does not contribute to the translation kinetic energy spectrum, the colloids' pairwise orbital translation injects energy into the system at values slightly smaller than  $q_\te{max}$. 
As more colloids are involved in the formation of larger vortices, the kinetic energy of the pair orbit is subsequently transferred to larger length scales and thus to smaller $q$ scales until a steady-state is reached. Since larger vortices store larger amounts of kinetic energy, the value of $E_q$ in Fig.~\ref{fig:Eq} grows for decreasing $q$. %

In the absence of substrate friction, the growth of $E_q$ towards smaller $q$ shown in Fig.~\ref{fig:Eq} continues until $q_{min}$, since the vortex size is only limited by the system size. On the other hand, the self-similar energy transport renders the energy spectrum to follow a power-law decay $q^{-5/3}$~\cite{mecke2023simultaneous}.  
For finite substrate friction, the energy spectrum attains a maximum at a wave number $q_{_E}$, as is indicated by the dotted vertical lines in Fig.~\ref{fig:Eq}. These maxima indicate the existence of a frictional cutoff at characteristic dissipation length scale $l_{_E} \simeq 2\pi/q_{_E}$~\cite{boffetta2012two}. 
Energy is injected by the activity of the rotors and then transported to larger scales until it is 
dissipated on scales where the attenuation by the substrate dominates. 
Accordingly, the self-similar energy transport from large to small $q$ and the power-law behaviour are also clearly disrupted. %
For large $q$, \ie, on short length scales, the substrate friction does not play a dominant role, such that $E_q$ does not depend much on $\gamma$. %
For finite values of $\gamma$, the colloid dynamics gets uncorrelated on inverse length scales $q \ll q_{_E}$. Hence, the energy spectrum there follows $E_q \propto q$ as can be inferred form Eq.~\eqref{eq:ennergy-spectrum} for a white, \ie, constant, power spectral density. %

The obtained measurements of $l_{_E}$ are related to the distances at which there is maximum storage of kinetic energy, and therefore to the average size of the vortices. 
The actual calculated values of $l_{_E}$ do not exactly correspond to the friction length $\lambda$, but they are both related by sharing the same functional dependence, as can be seen in the inset of Fig.~\ref{fig:Eq}.  
Similar to the case of the velocity correlation, the two values can be considered to be directly related by a correction factor $A_{_E}$, with $l_{_E} = A_{_E} \lambda$ and $A_{_E}=4.6$. 

\begin{figure*}[t!]
    \includegraphics[width=\textwidth]{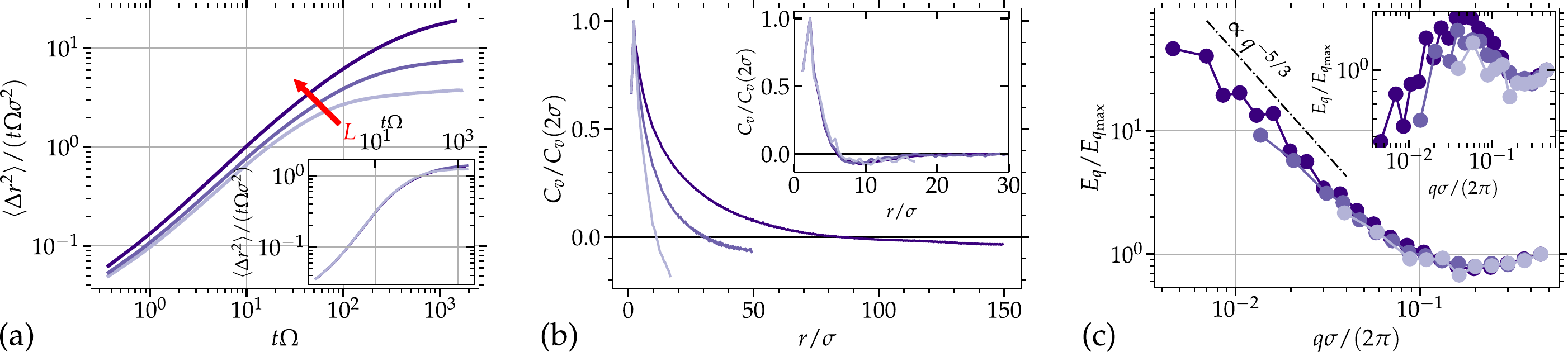}
	\caption{
        {\bf Finite size effects in the presence of substrate friction.}
        Simulation results for the collective behaviour of systems with different system sizes $L = 300\sigma$ (dark color lines),  $L=100\sigma$ (medium color lines), and   $L=35\sigma$ (light color lines). The main plots correspond to systems without friction,  $\gamma=0$, and the insets with friction, $(\gamma \tau) 10^3 = 4.7$.
        (a)~Time-normalised mean-square displacement $\ev{\Delta r^2}/t$, %
		(b)~velocity correlation function $C_v$, %
        and (c)~energy spectra $E_q$.
	}
	\label{fig:fse}
\end{figure*}

\bigskip
\noindent{\bf Finite-size effects. \ }
The consideration of the substrate friction is related to an intrinsic length scale in the system, such that is important to understand how it interferes with the system size, both in systems without and with substrate friction.  
For systems without or with negligible substrate friction, the size of the vortices is only limited by the size of the container or the simulation box, which also has a direct impact on the system dynamics. 
For simulations with $\gamma=0$ and various system sizes, mean-square displacement, velocity correlation function, and energy spectra are shown in Fig.~\ref{fig:fse}. 

The time-normalised mean-square displacement $\ev{\Delta r^2}/t$ in Fig.~\ref{fig:fse}a shows that the ballistic regime at short times displays an increase of the actuated velocity due to the influence of the vortex size. %
At longer times, the difference between the mean square displacement lines increases which indicates that not only the velocity but also the rotational diffusion time increases with system size.

In order to perform a comparative study of the velocity correlation function, we have to consider that the magnitude of the velocity depends on the system size, such that we normalise $C_v$ with the peak value $C_v(2\sigma)$. The results in Fig.~\ref{fig:fse}b are independent of the system size only at very short distances, since the dynamics provided by two or three rotors orbiting around each other is not affected much by the system size. 
The difference becomes much more prominent for slightly larger until very large distances, since the dominant minimum of $C_v$ in systems with $\gamma=0$ is located at a distance of $l_v=L/2$.

The energy spectra $E_q$ in Fig.~\ref{fig:fse}c shows a universal behavior for the full range of available $q$ values in each case. 
The $q^{-5/3}$ power-law emerging from the self-similar behavior extends also for the smallest $q$ scales, which validates that the largest vortex ranges over the whole system. %
In general, there are different options to normalise the power spectral densities. For simplicity, we choose the normalisation $E_q/E_{q_\te{max}}$ when comparing energy spectra of systems of different sizes, whereas spectra in Fig.~\ref{fig:Eq} where systems of equal size are compared, the normalization is done with the mean kinetic energy.

For systems with friction, the damping length $\lambda$ in Eq.~(\ref{eq:lambda}) arises as a relevant system length scale. 
As discussed before, the value of $\lambda$ limits the vortex size which affects most system properties. 
Simulation results with fixed  $\gamma \tau = 5 \times 10^{-3}$ and various system sizes are shown  as insets of the plots in Fig.~\ref{fig:fse}. 
Both,  $\ev{\Delta r^2}$ and $C_v$ in Fig.~\ref{fig:fse}a,~b show almost no difference related to the system size. 
This means that for this relatively large friction, the vortex size is not affected by the system size, and similarly the actuated velocity and rotational diffusion are not affected either. 
Similarly, the energy spectra $E_q$ in Fig.~\ref{fig:fse}c shows almost no dependence with the overall system size, in spite of differences related to the statistical accuracy of the data. 

\bigskip
\noindent{\bf Comparison with experiments. \ }
One relevant experimental example of these type of chiral systems is a suspension of rod-like colloidal particles with a perpendicular magnetic moment at one of their tips. 
Under the influence of an externally applied rotating magnetic field, the rods stand up and orient perpendicular to the substrate while rotating in sync with the external field developing the same type of vortex dynamics as described above~\cite{mecke2023simultaneous}.
Previous simulations were performed only in the absence of substrate friction. The measured actuated velocity in simulations was then a factor $4$ to $6$ times larger than in experiments~\cite{mecke2023simultaneous}, which can be easily related to the presence of  friction in the experimental system. 
For the measurements of the energy spectra, the simulation results were able to recover the $q^{-5/3}$ power-law dependence, but not the truncation of the spectra due to the limitation of the vortex size induced by the friction. 

\begin{figure}[!ht]
 	\includegraphics[width=0.48\textwidth]{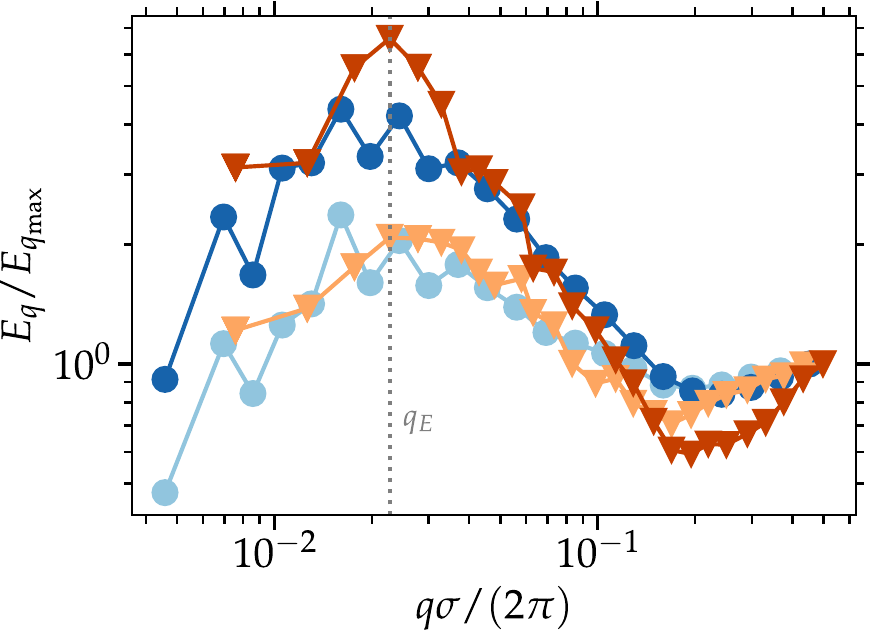}
	\caption{
        {\bf Comparison experimental and simulations measurements of the energy spectra.} 
		$E_q$ as obtained from experiments (triangles) and simulations (bullets) for two values of the rotor density, $\phi = 0.075$ (light colors) and $\phi = 0.144$ (dark colors). The vertical line indicates $q_{_E}$, the maximum of the spectra related to the dissipation length $l_{_E}$. Experimental data comes from Ref.~\cite{mecke2023simultaneous}.  Simulation data obtained in a system of size $L/\sigma=300$ and $\gamma = 0.05\Omega$, the last chosen to fit the experimental $q_{_E}$ value. %
	}
	\label{fig:Eq_comp}
\end{figure}
The question now is whether it is possible to match the value of the friction coefficient between simulation and experiments, such that the truncation of the spectra with both approaches quantitatively agrees. 
We consider the experimental energy spectra of the system shown in Fig.~\ref{fig:Eq}, measured for two values of the colloids packing fractions, $\phi=0.075$ and $\phi=0.144$. 
The truncation of the spectra occurs approximately at the same wave vector, $q_{_E}\sigma/(2\pi)\simeq 0.023$, for the two values of the  density which is related to a single friction coefficient, given by same experimental conditions in both measurements. 
Simulations are then performed at the same densities as in the experiments, first with a value of the friction coefficient directly estimated from the above value of $q_{_E}$ and calculated from the related $l_{_E}=2\pi/q_{_E}=A_{_E}\lambda$,  which by using Eqs.~(\ref{eq:lambda}) and.~(\ref{eq:gamma-short}) results in density of virtual particles $\rhos=0.00012$ in the MPC simulations, which corresponding to a friction value $\gamma=0.03\Omega$ in this system.   
This estimate provides satisfactory results, but with a slight underestimation of the truncation length $q_{_E}$, due to the difficulties in the determination of $A_{_E}$. Accordingly, we adjust to $\rhos=0.0002$, corresponding to $\gamma=0.05\Omega$. 
Figure~\ref{fig:Eq_comp} then displays a very good agreement using the adjusted values, which furthermore very nicely proves the emergence of a relevant length scale in the system due to the presence of a substrate friction.  

\section{\label{sec:summary}Discusssion}
Solutions of particles sedimented on a surface, or with motions restricted to a thin fluid layer, are frequently investigated with two-dimensional theoretical approaches.  Hydrodynamic interactions might though lead to unphysical behaviour due to the so-called Stokes paradox, which is related with the too slowly decaying fluid flows in two dimensions. This has been ignored or circumvented by the consideration of three-dimensional setups with a significantly increased analytical or computational effort.
Here, we propose an efficient method to incorporate the effect of the friction occurring at the solid substrate or at the interfaces for a particle-based mesoscopic model known as multiparticle collision dynamics. The idea is to account for the momentum interchange with a reservoir by interactions with a very small number of virtual particles.
The proposed method correctly reproduces the low Reynolds number flow behaviour of the Brinkman equation, \ie, Stokes equation with an additional friction term given by a well-defined substrate friction coefficient $\gamma$.
The friction leads to a decay of the created flows on a typical damping length scale $\lambda$, which can be analytically related to the fluid properties.
The introduction of substrate friction leads to a frictional cutoff of the created flows and thus serves as a regularisation of the hydrodynamics in two dimensional systems, which additionally makes most results independent on the size of the simulated system. %

Then, a chiral active system consisting of a suspension of rotating colloids sedimented on a solid surface is considered. Vortices of different sizes are dynamically formed as a result of the colloids individually induced flow fields which drag other colloids exerting mutual rotational stresses. 
The dynamics exhibits active turbulence with a self-similar behaviour, although 
experimentally obtained power spectra exhibit a cutoff at short wavelengths, indicating the emergence of a length scale, at which most of the kinetic energy is stored~\cite{bililign2022motile,mecke2023simultaneous}, which limits the actual maximum vortex size.
Here we show how the presence of substrate friction imposes a dominant vortex or cutoff scale such that the maximal vortex size is on the order of the solvent damping length $\lambda$, which indirectly also affects other properties such as the average actuated particle velocity, or the rotational diffusion coefficient.   %
The estimation of the vortex size is  done with two quantitative approaches. 
From the velocity spatial correlation function $l_v$ is obtained, which relates to the average centre-to-centre distance of neighbouring vortices with opposite vorticity.
From the energy spectra $l_{_E}$ is obtained. This relates to the distances where the storage of kinetic energy is maximum, which corresponds to the typical length scale of the dominating vortices. 
A third estimation could have been obtained by a quantitative image-analysis, but a qualitative  inspection of the vorticity fields in Fig.~\ref{fig:vorticity-density-collective} already shows that the value would have the same trend and similar values than $l_v$ and  $l_{_E}$.
These lengths are evaluating different properties, all related to the properties of the colloids dragged by the solvent, such that they scale in the same way with the solvent damping length $\lambda$.
This behaviour convincingly shows that the vortex size decreases with an increasing damping length provided by the substrate friction.
In particular, additional simulations are performed to precisely agree with the experimental power spectra, characterizing the wavelength at which is the power law scaling is cutoff. 
The substrate friction can therefore be used as tuning parameter to manipulate the distribution of kinetic energy over length scales. %

For active chiral systems, the effect of substrate friction can be used to tune the system properties, and is expected to translate into further effects. A system of rotating colloids in circular confinement has for example experimentally shown to extend the range of the emerging edge current with decreasing friction~\cite{soni2019odd}. 
Numerous other systems in quasi-two-dimensional confinement, such as bacterial motion or surface effects will also profit from the consideration of a tunable surface friction which will extend the applicability of the approach. 
Finally, we expect that the proposed extension of the method can be also adapted to systems on three dimensions and also to other methods such Dissipative Particle Dynamics, or Lattice Boltzmann simulations which will be useful for a large number of applications.

\section{Methods}
\begin{footnotesize}

\bigskip
\noindent{\bf Multiparticle collision dynamics. \ }
The basic simulation method we employ is a two-dimensional variant of the mesoscopic MPC algorithm, which includes thermal fluctuations and hydrodynamic interactions and accordingly reproduces low-Reynolds number flow efficiently if suitable parameters are employed~\cite{ripoll2005dynamic}. %
The fluid consists of point particles, their dynamics is governed by and alternating streaming and collision steps.
In the streaming step, the particle positions $\bm{r}_i$  are updated by employing ballistic transport at velocities $\bm{v}_i$ for the duration of the collision time $h$,
\begin{align}
    \bm{r}_i(t+h) = \bm{r}_i(t) + \bm{v}_i(t) h\,.
\end{align}
The fluid consists of point particles, their dynamics is governed by and alternating streaming and collision steps.
In the subsequent collision step, the particles are sorted into square collision cells of length $a$. Then particles exchange momentum by which each fluid particle's relative velocity with respect to the centre-of-mass velocity of the given collision cell is rotated by an angle of $\pm\vartheta$, with equal probability, \ie,
\begin{align}
\label{eq:collision}
    \bm{v}_i(t+h) = & \bm{v}_\zeta(t)  + 
    \begin{pmatrix}
    \cos\vartheta & -\sin\vartheta\\
    \sin\vartheta & \cos\vartheta
  \end{pmatrix}
    \cdot (\bm{v}_i(t) - \bm{v}_\zeta(t)) \nonumber \\ 
 & -(\bm{I}_\zeta^{-1}\cdot \Delta \bm{L}_\zeta) \times (\bm{r}_i - \bm{r}_\zeta)
\end{align}
where  $\bm{r}_\zeta$, $\bm{v}_\zeta$, $\bm{I}_\zeta$, and $\Delta \bm{L}_\zeta$  refer to the centre-of-mass position, velocity, moment of inertia, and change in angular momentum in a given collision cell, respectively. %
The first two terms in Eq.~(\ref{eq:collision}) provide the local conservation of mass and linear momentum, which is sufficient to ensure the proper hydrodynamic behavior~\cite{malevanets2000}. To also account for local angular momentum conservation, here we consider the additional third term in Eq.~(\ref{eq:collision})~\cite{noguchi2008transport}. This has shown to avoid the occurrence of unphysical torques in the fluid when studying rotational flows~\cite{pooley2005kinetic}. 
Random shift of the collision cell grid before each collision is used in order to ensure that Galilean invariance is retained and that fluid particles are not strongly correlated to their direct neighbours~\cite{ihle2001stochastic,gompper2009multi}.
In systems with a continuous energy input, such as active  systems, we employ a cell level thermostat~\cite{huang2010cell}, which guarantees an average constant system temperature $\kBT$. %
We employ an average fluid density of $\rho=10 m/a^2$, a rotation angle $\vartheta=\pi/2$, and collision time $h = 0.02a\sqrt{m/(\kBT)}$, resulting in a viscosity of $\eta = 17.9 \sqrt{m\kBT}/a$~\cite{noguchi2008transport}. We consider $m=a=\kBT=1$ as the system units, such that time is measured in units of $\tau = a\sqrt{m/(\kBT)}$.

\bigskip
\noindent{\bf Analytical estimation of MPC friction coefficient. \ }
We now explicitly relate the presence of a certain density of virtual particles $\rhos$ with the induced friction coefficient $\gamma$ in a MPC fluid.    
The friction can be estimated as  the average amount of dissipated momentum in an arbitrary direction (here  $x$) during collision time $h$, normalised by the total cell momentum, such that  
\begin{align} \label{eq:gamma1}
    \gamma = -\ev{ \frac{\sum_{i\in\zeta} v_i^{(x)}(t + h) -\sum_{i\in\zeta} v_i^{(x)}(t)}{ h \sum_{i\in\zeta} v_i^{(x)}(t)} }\,.
\end{align}
By construction, in the absence of virtual particles,  Eq.~\eqref{eq:collision} exactly conserves momentum
and $\gamma = 0$. 
To calculate this average, we consider the collision in Eq.~\eqref{eq:collision} together with
a uniform distribution of fluid and virtual particles in the collision cell, and average over the rotation matrix, with the assumption of molecular-chaos, \ie, that the velocities of different particles in the same cell are completely uncorrelated~\cite{ripoll2005dynamic, pooley2005kinetic,noguchi2008transport}. 
Note that when considering virtual particles in the calculation of Eq.~(\ref{eq:collision}), the velocities and the randomly chosen position of the virtual particles are considered for the calculation of the values of  $\bm{r}_\zeta$, $\bm{I}_\zeta$, and $\Delta \bm{L}_\zeta$ in Eq.~(\ref{eq:collision}). In principle the conservation of angular momentum in these cells is not critical, but it is here included both, for consistence and for simplicity.

Since we focus in the two-dimensional case, the rotational matrix considers both possible directions of rotation, such that the average in Eq.~(\ref{eq:gamma1}) is,
\begin{align}
\label{eq:gamma-long}
        \gamma = \frac{1}{h} \frac{\rhos}{\rho + \rhos} \left[ 1- \cos\vartheta - \frac{1}{2(\rho - 1)} (1 + \cos\vartheta) \right]\,.
\end{align}
The two first terms inside the brackets correspond to the first terms in Eq.~(\ref{eq:collision}) related to the conservation of mass and linear momentum, the third term inside the brackets corresponds to the third term in Eq.~(\ref{eq:collision}) related to the conservation of angular momentum.
 
\bigskip
\noindent{\bf The Brinkman equation. \ }
The hydrodynamic equations of motion of a fluid in the low Reynolds regime is generally described by the well-known Stokes equation.
When friction plays an important role, the Stokes equation needs an additional term that accounts for a constant rate of momentum absorption~\cite{boffetta2012two, banerjee2017odd, liu2020oscillating}, which is directly proportional to the fluid velocity $\bm{u}$ and the substrate friction coefficient $\gamma$, so that
\begin{align}
	\label{eq:stokes-friction}
	\frac{1}{\rho} \partial_\alpha p = \nu \partial_\beta \partial_\beta u_\alpha - \gamma u_\alpha\,,
\end{align}
with $\alpha, \beta = x, y$, closure $\partial_\alpha u_\alpha = 0$, and summation over repeating indices is implied.
This equation is known as the {\em Brinkman equation}, and is also known from low-Reynolds flow in porous media~\cite{koplik1983viscosity}.
A check of the validity of this equation is therefore an important test to demonstrate the soundness of our approach.

When considering a system with radial symmetry such as the one describe in the main text for the circular Couette flow, the velocity can be taken as simply radial dependent $\bm{u} = u_\varphi(r) \hat{e}_\varphi$. The pressure term is also radial-dependent such that it has no contribution in the tangential direction. The Brinkman equation~\eqref{eq:stokes-friction} in polar coordinates is then
\begin{align}
	\label{eq:fluid-friction}
	0 = \partial_r^2 u_\varphi + \frac{1}{r} \partial_r u_\varphi - \frac{1}{r^2} u_\varphi - \frac{\gamma}{\nu} u_\varphi\,.
\end{align}
Substituting $\widetilde{r} \equiv i r/\lambda$, with $i$ is the imaginary unit, transforms the equation into
\begin{align}
	0 = \widetilde{r}^2 \partial_{\widetilde{r}}^2 u_\varphi + \widetilde{r} \partial_{\widetilde{r}} u_\varphi + (\widetilde{r}^2 - 1) u_\varphi\,.
\end{align}
This is known as the Bessel equation, and the general solution is 
\begin{align}
	u_\varphi(r) = a J_1(\widetilde{r}) + b Y_1(\widetilde{r}), 
\end{align}
with $J_1(\widetilde{r})$ and $Y_1(\widetilde{r})$ the Bessel functions of the first and second kind, respectively~\cite{akhmedova2019selected}, and the integration constants $a$ and $b$ to be determined by the boundary conditions. 
$u_\varphi(R_\te{i}) = R_\te{i}\Omega$ and $u_\varphi(R_\te{o}) = 0$. %
\end{footnotesize}

\subsubsection*{Acknowledgments}
J.M. gratefully acknowledges the National Natural Science Foundation of China for supporting this work within the Research Fund for International Young Scientists under grant number 12350410368. J.M. and M.R. gratefully acknowledge the Gauss Centre for Supercomputing e.V. (www.gauss-centre.eu) for funding this project by providing computing time through the John von Neumann Institute for Computing (NIC) on the GCS Supercomputer JUWELS at Jülich Super-computing Centre (JSC) and the Helmholtz Data Federation (HDF) for funding this work by providing services and computing time on the HDF Cloud cluster at the Jülich Supercomputing Centre (JSC). Y.G. acknowledges funding support from the Natural Science Foundation of Guangdong
Province [2024A1515011343].

\subsubsection*{Data availability}
The data that support the findings of this study are available from the corresponding
authors upon reasonable request.

\subsubsection*{Code availability} 
The custom code for the simulations on GPUs is available from the corresponding authors
upon reasonable request.

\subsubsection*{Author contributions}
J.M., G.G and M.R. designed the implementation of substrate friction.
J.M. and M.R. designed the investigation.
J.M. wrote the simulation code, conducted the numerical simulations, and performed the analytical calculations.
Y.G. designed the experimental setup and performed the experiments. 
J.M., Y.G., and M.R. discussed the results.
J.M. and M.R. wrote the original draft.
All authors finalised and approved the manuscript. 

\bibliographystyle{naturemag}
\bibliography{Bibliography,bib_mr}

\end{document}